\documentclass[prl,superscriptaddress, twocolumn, amsmath,amssymb
]{revtex4-2}

\usepackage{pifont} 

\usepackage[utf8]{inputenc}
\usepackage{bibunits}

\usepackage{setspace}
\usepackage[T1]{fontenc}
\usepackage{hyperref}
\usepackage{multirow}
\usepackage{ulem}
\usepackage{manfnt}

\usepackage{graphicx}
\usepackage{subfigure}
\usepackage{float}
\usepackage{xcolor}

\usepackage{amsmath}
\usepackage{amsfonts, relsize, color}
\usepackage{mathrsfs}
\usepackage{mathtools}
\usepackage{physics}
\usepackage{bbm}
\usepackage{bbold}

\newcommand{\ncmd}{\newcommand}
\newcommand{\beq}{\begin{equation}}
\newcommand{\eeq}{\end{equation}}
\ncmd{\nn}{\nonumber}
\ncmd{\mbf}[1]{\bs{#1}}
\ncmd{\gam}{\gamma}
\ncmd{\lam}{\lambda}
\ncmd{\kap}{\kappa}
\ncmd{\Lam}{\Lambda}
\ncmd{\Gam}{\Gamma}
\ncmd{\Dl}{\Delta}
\ncmd{\Ups}{\Upsilon}
\ncmd{\Om}{\Omega}
\ncmd{\eps}{\epsilon}
\ncmd{\veps}{\varepsilon}
\ncmd{\vphi}{\varphi}
\ncmd{\vtheta}{\vartheta}
\ncmd{\tw}{\text{w}}
\ncmd{\pd}{\partial}
\ncmd{\pll}{\parallel}
\ncmd{\mc}{\mathcal}
\ncmd{\mf}{\mathfrak}
\ncmd{\bs}{\boldsymbol}
\ncmd{\half}{\frac{1}{2}}
\ncmd{\tilJ}{\tilde{J}}
\ncmd{\avg}[1]{\langle{#1}\rangle}
\ncmd{\note}[1]{{\color{red}{\ding{168} [#1]}}}
\ncmd{\eq}[1]{Eq. \eqref{#1}}
\ncmd{\fig}[1]{Fig. \ref{#1}}
\ncmd{\suppl}{\note{`Supplementary Information'}}
\ncmd{\todo}[1]{{\color{magenta} \textit{\ldots #1}}}
\ncmd{\psicl}{\psi_\text{cl}}
\ncmd{\XC}{\text{XC}}
\ncmd{\an}[1]{{\color{blue}#1}}
\newcommand*\cube{\mbox{\mancube}}

\begin{document}

\title{Realizing fracton order from long-range quantum entanglement in programmable Rydberg atom arrays}

\author{Andriy H. Nevidomskyy}
 \email{nevidomskyy@rice.edu}
\affiliation{Department of Physics and Astronomy, Rice University, Houston, TX 77005, USA}
\affiliation{Rice Center for Quantum Materials, Smalley--Curl Institute, Rice University, Houston, TX 77005, USA}

\author{Hannes Bernien}
\affiliation{Pritzker School of Molecular Engineering, University of Chicago, Chicago, IL 60637, USA}

\author{Alexander Canright}
\affiliation{Department of Physics and Astronomy, Rice University, Houston, TX 77005, USA}

\date{\today}


\maketitle




\onecolumngrid
\noindent
\textbf{Storing quantum information, unlike information in a classical computer, requires battling quantum decoherence, which results in a loss of information over time. 
To achieve error-resistant quantum memory, one would like to store the information in a quantum superposition of degenerate states engineered in such a way that local sources of noise cannot change one state into another, thus preventing quantum decoherence. One promising concept is that of fracton order -- a phase of matter with a large  ground state degeneracy that grows subextensively with the system size. Unfortunately, the models realizing fractons are not friendly to experimental implementations as they require unnatural interactions between a substantial number (of the order of ten) of qubits. We demonstrate how this limitation can be circumvented by leveraging the long-range quantum entanglement created using only pairwise interactions between the code and ancilla qubits, realizable in programmable tweezer arrays of Rydberg atoms. We show that this platform also allows to detect and correct certain types of errors \textit{en route} to the goal of true error-resistant quantum memory.} 
\vspace{8mm}
\twocolumngrid

\noindent

One approach to battle the quantum decoherence is via active quantum error correction. 
However, this approach typically requires large redundancy of auxiliary (ancilla) qubits compared to the number of logical ``code qubits'' and a high fidelity of quantum gates is necessary to assure the effectiveness of error correction~\cite{shor,knill1998}.  
Recently, a different set of ideas has emerged, where the protection against quantum information loss is predicated not on error correction protocols, but rather on the inability of the information to dissipate because the quantum memory is stored in degenerate ground states that are stable against action by local operators. 
In the past decade, a set of models have been constructed based on so-called fracton matter~\cite{Haah2011,Vijay2015,Vijay2016},  which hinges on the existence of  immobile quasiparticles, called \textit{fractons}, which enable a subextensive ground state degeneracy that grows exponentially in the linear system size. 
In some fracton systems, these phenomena are associated with desirable (though not ideal) 
self-correction properties~\cite{bravyi-haah2013}, minimizing the need for active error correction.
It is thus very desirable to realize fracton matter in a laboratory, however 
the principal challenge is that all known  lattice models, such as the X-cube code~\cite{Vijay2016}, require `unnatural' interactions between many (up to 12) particles at a time. 

Here we propose a concrete protocol for realizing the ground state of the fractonic X-cube model using programmable Rydberg tweezer arrays as a platform,  exploiting the recently developed dual-species architecture in which the two atomic species can be measured and controlled without crosstalk~\cite{Singh2023, anand_dual-species2024}. 
Instead of directly implementing the stabilizer terms of the X-cube model, which is challenging because of the multi-atom nature of the necessary interactions, our proposal requires only two-qubit gates between the ancilla and code qubits, whose implementation is readily available and can be performed efficiently and in parallel using programmable tweezer arrays~\cite{levine_gates2019,evered_parallel-gates2023}. 
We further demonstrate that our scheme -- although not routed in the Hamiltonian approach, and therefore not readily designed to address the excited states of the X-cube code -- nevertheless allows us to detect and correct the errors that correspond to the fracton and lineon excitations of the X-cube model, with associated  subdimensional mobility. 

\vspace{2mm}
\noindent
\textbf{Programmable Rydberg tweezer arrays.}
Neutral atoms trapped in optical tweezer arrays have become a front-runner for both analog quantum simulation~\cite{browaeys_review2020,scholl_simulation2021,Ryd-256qubit-ebadi-lukin2021} and digital quantum computation~\cite{kaufman_review2021,graham_algorithms2022,Ma2023,bluvstein_logical-processor2024}. By dynamically controlling the position of the optical tweezers arbitrary arrays of atoms can be created in one dimension~\cite{Ryd-chain-endres2016}, two dimensions~\cite{Barredo2016,Kim2016} and even three dimensions~\cite{Barredo2018, Lee2016}. Coherent interactions between the atoms are then generated by coupling to highly excited Rydberg states~\cite{Ryd-review-saffman2010}. 
The combination of flexible geometries with  long-range interactions between atoms was instrumental to the recent implementation~\cite{toric-Ryd-semeghini2021} of the toric code Hamiltonian, whose ground state possesses long-range entanglement~\cite{Kitaev-toric}.

An alternative to the analog approach for realizing long-range entanglement, is to use a gate-based approach as was recently done to realize logical qubit states encoded in multiple entangled atomic qubits~\cite{bluvstein_logical-processor2024}. Here, we propose to follow such a digital strategy to realize fracton order in a dual-species Rydberg array. 


\vspace{2mm}
\noindent
\textbf{Creating the cluster state.} 
The cluster state is a highly entangled multiqubit state that has two key properties~\cite{briegel-raussendorf2001}: (i) it is \textit{maximally connected}, meaning that any two qubits can be projected into a pure Bell state by local measurements on a subset of other qubits, and (ii) it has high \textit{persistence of entanglement}  (in fact, scales with the number of qubits $N$) -- defined as the minimum number of local measurements necessary to completely disentangle the state. 
The latter property is particularly important as it protects the robustness of entanglement -- in contrast with, for instance, Greenberger--Horne--Zeilinger (GHZ) states which can be reduced to a product state by a single local measurement.

\begin{figure}[tbh]
   \subfigure[]{
    \centering
  \includegraphics[width=0.45\textwidth]{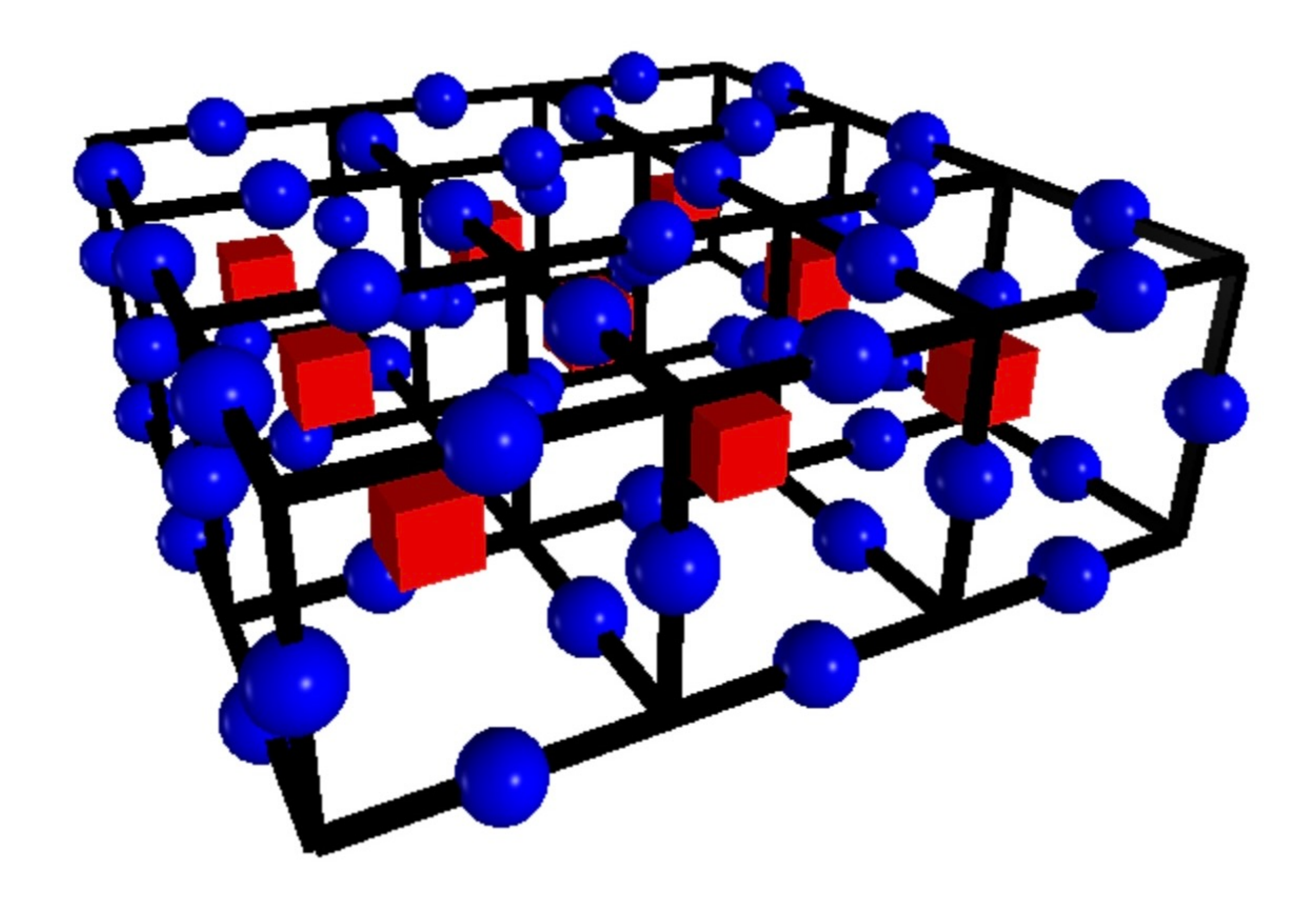}
  } \\
   \subfigure[]{
  \includegraphics[width=0.2\textwidth]{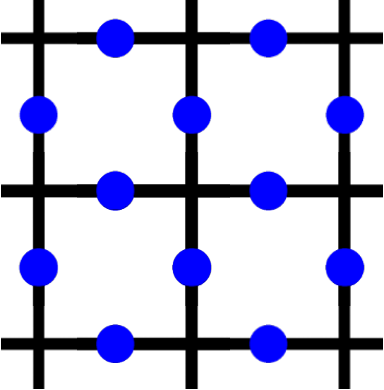}
  }
   \subfigure[]{
  \includegraphics[width=0.2\textwidth]{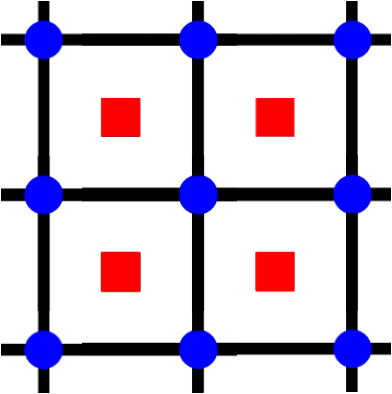}
  } \hspace{1mm}
  \caption{(a) A Rydberg atom array composed of two species: the code qubits (blue) and ancilla qubits (red). This is achieved using alternating layers (b), (c) with the code qubits occupying the vertices (c) and the bond centers (b) of the square lattice, while the ancilla qubits inhabit the cube centers.}
  \label{fig:lattice}
\end{figure}

\begin{figure}[b!]
    \subfigure[]{
   \includegraphics[width=0.22\textwidth]{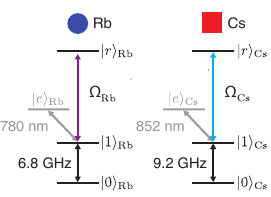}
  } \hspace{1mm}
  \subfigure[]{
  \includegraphics[width=0.20\textwidth]{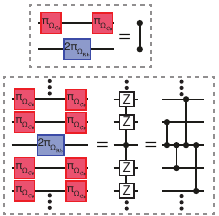}
  }\\
  \subfigure[]{
  \includegraphics[width=0.4\textwidth]{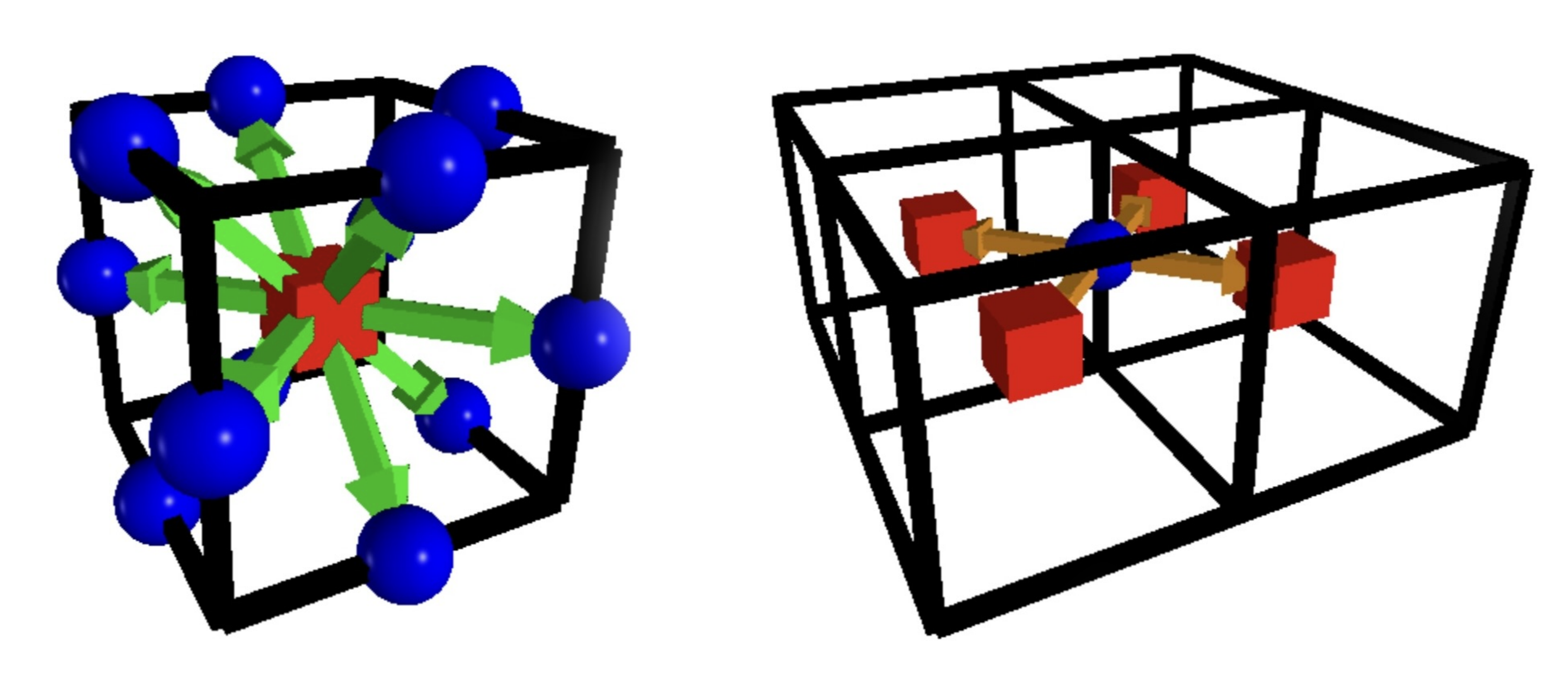}
  } 
  \caption{(a) The level structure of rubidium (Rb) and cesium (Cs) atoms excited to their respective Rydberg states via species-selective two-photon transitions. The qubits are encoded in the hyperfine $\ket{0} \leftrightarrow \ket{1}$   manifold dressed by the Rydberg $\ket{r}$ states, with $\ket{0}_\text{Rb} = \ket{F = 1, mF = 0}, \ket{1}_\text{Rb} = \ket{2, 0}, \ket{0}_\text{Cs} = \ket{3, 0}, \ket{1}_\text{Cs} = \ket{4, 0}$. Qubit readout is performed via fluorescence on the $\ket{1} \leftrightarrow \ket{e}$  transitions. (b) The interspecies CZ gate is obtained, up to single-qubit phases, by performing $\pi$-$2\pi$-$\pi$ sequence of pulses between the hyperfine ($\ket{0},\ket{1}$) states~\cite{Jaksch2000}.  This results in a two-qubit gate for two adjacent atoms. In a configuration where the ancilla qubit interacts with the code qubits but the code qubits do not interact with each other, the same pulse sequence results in a multi-target controlled Z  (CZ$_{12}$) gate. (c) The CZ gates (green arrows) are applied  between the ancilla and each of the 12 neighboring code qubits by moving the tweezer arrays holding the ancillae (Cs) relative to the other atoms (Rb). Each code qubit is connected to four ancilla qubits positioned in the centers of adjacent cubes.}
  \label{fig:cluster}
\end{figure}

The simplest way to define the cluster state is by specifying the set of its stabilizing generators. For this, we first need to describe the lattice geometry, which we choose to be in the form of a cubic lattice with the code qubits (blue in Fig.~\ref{fig:lattice}) occupying the bond centers. The other component are the ancilla qubits (red), which we place on the dual lattice, i.e. in the centers of the cubes as depicted in Fig.~\ref{fig:lattice}.
Next, we define the cluster operators centered on the code-qubit, $C_c$ and their analog centered on the ancilla qubits $C_a$ of the form
\begin{align}
C_c = X_c \prod_{a\in \partial c} Z_a, \quad
C_a = X_a \prod_{c\in \partial a} Z_c,   \label{eq:cluster-ops} 
\end{align}
where the product is over the ancilla (code) qubits adjacent to the given code (ancilla) qubit, as shown in Fig.~\ref{fig:cluster}(c). 
The cluster state is then defined as a state that is simultaneously an eigenstate (with eigenvalue, say, +1) of all the $C_c$ and $C_a$ cluster operators:
$C_c\ket{\psicl} = \ket{\psicl} = C_a \ket{\psicl}$. 
We note that this is a three-dimensional generalization of the checkerboard lattice of qubits and ancillae considered in Refs.~\cite{raussendorf-bravyi-harrington2005,raussendorf-harrington2007}.

To prepare the cluster state, we first initialize both the code and ancilla qubits in the $\ket{+}=\frac{1}{\sqrt{2}}(\ket{1} + \ket{0})$ eigenstate of the $X$ operators. 
We then apply a two-gate control-Z (CZ) operator on each pair of adjacent ancilla and code qubits -- this ensures that the resulting state $|{\tilde{\psi}_\text{cl}}\rangle$ is an eigenstate, with eigenvalue +1 (see \textit{Methods}), of both the $C_a$ and $C_c$ operators appearing in Eq.~\eqref{eq:cluster-ops}. 
%

Crucially, the recent advances in programmable tweezer arrays make the implementation of the aforementioned CZ operators feasible, and achievable in parallel. Specifically, qubit states are encoded in long-lived hyperfine states with coherence times that can exceed seconds~\cite{Manetsch2024}. By coupling two adjacent atoms that are sufficiently close to each other to the Rydberg state and sending a certain sequence of pulses shown in Fig.~\ref{fig:cluster}(b), a two qubit gate can be realized. The principle relies on Rydberg blockade, in which a conditional phase is picked up by one atom dependent on the state of the other resulting in a CZ gate~\cite{Jaksch2000}, see \textit{Methods}. Recently these gates have been implemented with high fidelities~\cite{evered_parallel-gates2023,Ma2023}.

In order to achieve the 12 CZ gates between the ancilla qubit and the adjacent code qubits we propose two complementary strategies. 
The first relies on the coherent transport of the atoms that allows one to change the connectivity of the qubits between operations~\cite{bluvstein_quantum-processor2022}. 
By then moving the ancilla qubit sequentially to each of the adjacent code qubits, as indicated by the green arrows in Fig.~\ref{fig:cluster}(c) and performing the CZ gate on the ancilla-code qubit pair, the desired cluster state is generated. Importantly, these operations are performed in all elementary cubes in parallel. 

The second strategy to efficiently implement the 12 CZ gates (between 12 code qubits and one ancilla), uses the unique properties of dual-species Rydberg arrays that can achieve the desired connectivity without movement of the atoms. By using F\"orster resonances between suitably chosen Rydberg states, the interactions between ancilla and code qubits can be strongly enhanced over the interactions of the same type~\cite{Beterov2015,anand_dual-species2024}. This allows us to arrange the atoms such that the ancilla qubit experiences strong Rydberg blockade with the adjacent 12 code qubits, while the code qubits among themselves will not be blockaded. The coupling to the Rydberg states therefore leads to the central ancilla qubit controlling the phase that is being picked up on each adjacent code qubit without the code qubits interacting with one other. This results in a CZ${}_{12}$ gate which is equivalent to 12 CZ gates between the ancilla and each code qubit (for more details see \textit{Methods}). The advantages of this dual-species implementation is that there is no need for atomic movement, which is slow compared to the gate speed, and that there will be less time spent in Rydberg states, which is a main source for gate infidelities.

\vspace{2mm}
\noindent
\textbf{Creating the eigenstate of X-cube.}  The cluster state thus prepared has a highly persistent (in the sense defined above) multi-qubit entanglement, which is a resource for quantum computing. It was realized early on that this resource can be leveraged to achieve highly nontrivial quantum states, such as the toric code, on the code qubits by performing selective measurements on ancilla qubits only~\cite{raussendorf-bravyi-harrington2005,aguado-ancillas-create-anyons2008,piroli-measurements2021}. Crucially, this allows the preparation of long-range entangled states, such as the toric code, with a finite-depth quantum circuit (augmented by measurements on the ancillae).

\begin{figure}[tbh]
 \subfigure[]{
    \centering
    \includegraphics[width=0.25\textwidth]{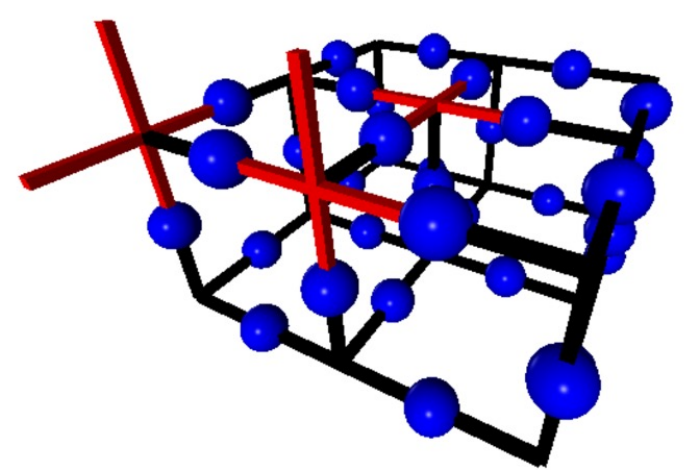}
  } 
 \subfigure[]{
    \centering
    \includegraphics[width=0.16\textwidth]{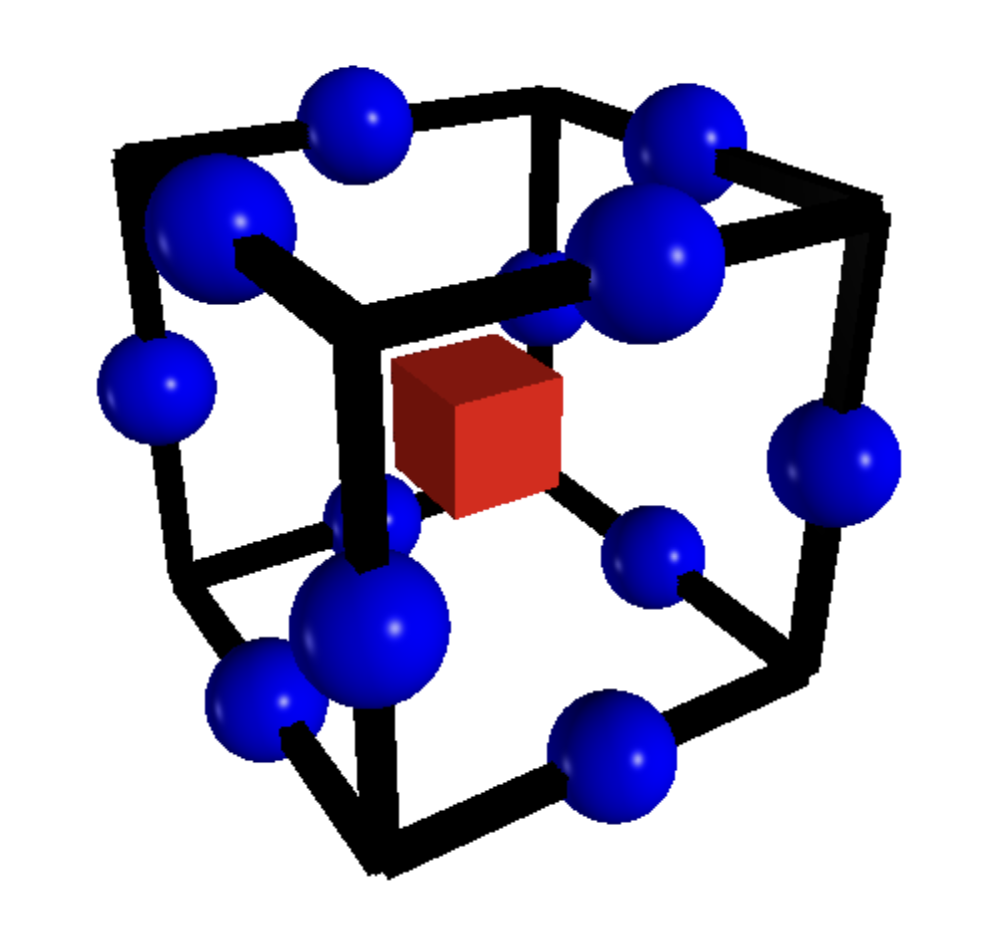}
  } 
\\  
    \subfigure[]{
    \centering
    \includegraphics[width=0.33\textwidth]{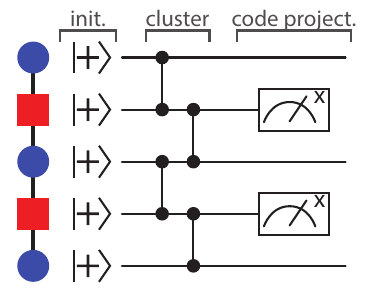}
  } 
  \caption{(a) Three kinds of the star operators appearing in the X-cube model Eq.~\eqref{eq:Xcube}. (b) Measuring the ancilla (red) in the $X$-basis with eigenvalue $m_a=\pm1$, enforces, through the property of the cluster state, the cube operator $B_\text{cube}(a)$ to have the same eigenvalue. (c) Section of the equivalent quantum circuit to generate the $\ket{\tilde{\XC}}$ state, starting from the $\bigotimes_{a,c}\ket{+}$ state on both the code and ancilla qubit, where $\ket{+}= \left(\ket{1}+\ket{0}\right)/\sqrt{2}$. The subsequent measurement of all the ancillae projects the cluster state onto the desired 
  state of the code qubits. }
    \label{fig:circuit}
\end{figure}

Here, we show how one can prepare the ground state of the X-cube model using a similar strategy. Recall that the X-cube model is defined on the edges of the  cubic  lattice  as a sum of commuting stabilizers \mbox{$H_X = - \sum A_\text{star} - \sum B_\text{cube}$} defined on code qubits as follows
\begin{align}
A_\text{star} &= \prod_{c\in s_{xy}} X_c +  \prod_{c\in s_{xz}} X_c +  \prod_{c\in s_{yz}} X_c, \nonumber\\ 
B_\text{cube} &= \prod_{c\in \cube} Z_c,
\label{eq:Xcube}
\end{align}
where each of the star operators includes the four  qubits in the $\{xy,xz,yz\}$ plane, respectively, shown in Fig.~\ref{fig:circuit}(a) and $B_\text{cube}$ is a product of 12 qubits on the cube's edges, Fig.~\ref{fig:circuit}(b). 
The ground state of the X-cube model $\ket{\XC}$ satisfies $A_\text{star}\ket{\XC} = +1\ket{\XC}$ and $B_\text{cube}\ket{\XC} = +1\ket{\XC}$ for all star and cube operators. Written explicitly (denoting the set of all cubes by $C$)
\beq 
\label{eq:XC}
\ket{\XC} \propto \prod\limits_{a\in C}\left(\mathbb{1} + B_\text{cube}(a) \right) \ket{+}^{\otimes M},
\eeq
where $M$ is the total number of code qubits. 

By construction of the cluster state $\ket{\psicl}$ in Eq.~\eqref{eq:cluster-ops}, it is an eigenstate of all the $X_c$ operators, and therefore already satisfies $A_\text{star}\ket{\psicl}= +1\ket{\psicl}$, similar to the star operators in the 2D toric code~\cite{bolt-foliated-ECC2016}. 
We now show how to obtain an eigenstate of the X-cube model $\ket{\XC}$, using projective measurements on the ancilla qubits. Indeed, measuring a given ancilla in the $X$ basis projects the state onto the definite eigenvalue $m_a=\pm 1$ of the ancilla $X_a$ operator. Since $\ket{\psicl}$ is an eigenstate of the cluster operator $C_a$ in Eq.~\eqref{eq:cluster-ops}, it follows that the resulting state, after the measurement, is an eigenstate (with eigenvalue $m_a$) of the cube operator defined around the ancilla $B_\text{cube}(a) =\prod_{c\in\partial a}Z_c$. Upon performing measurements on all the ancillae, we thus obtain the state
$\ket{\tilde{\XC}} \propto \prod\limits_{a}\left[\mathbb{1} + m_a B_\text{cube}(a) \right] \ket{+}^{\otimes M}$. 
We note that the CZ state can be equivalently represented as a combination of the single-qubit Hadamard gate and the CNOT gate, with the equivalent quantum circuit shown schematically in Fig.~\ref{fig:circuit-CX} (for more details see methods). We note that this quantum circuit is similar in spirit to the so-called dynamic quantum circuits, in which measurements on the ancillae (possibly combined with classical feed-foward) were used to ``teleport'' long-range quantum gates across a locally connected bus of ancillae~\cite{baumer_dynamic-circuits2023} and to create non-local entanglement between code qubits, for instance the GHZ state~\cite{piroli-measurements2021}.

The final step is to convert the state $\ket{\tilde{\XC}}$ to the desired X-cube eigenstate Eq.~\eqref{eq:XC}. Realizing that the product of all cube operators
\beq
\label{eq:cube-product}
\prod_a B_\text{cube}(a)  = \mathbb{1}
\eeq 
implies that the product of all measured eigenvalues $\prod_a  m_a = 1$. Given a set of $m_a$ satisfying this condition, it is always possible to find such a product of $X$ operators (denoted by $X_m$) that $X_m(\mathbb{1} + m_a B_\text{cube}(a)) X_m = (\mathbb{1} + B_\text{cube}(a))$, $\forall a$, similar to the construction of the 2D toric code out of the cluster state~\cite{raussendorf-bravyi-harrington2005,piroli-measurements2021}. By applying the product of such local unitaries $X_m$ thus converts $|\tilde{\XC}\rangle \longmapsto \ket{\XC}$.

We now turn to the practical issue of implementing this program in a tweezer array of Rydberg atoms. Given the technical challenges associated with creating truly 3-dimensional individually addressable arrays, we propose to focus as a first step on the three-layer quasi-2D network of code qubits and ancillae  shown in Fig.~\ref{fig:lattice}. Experimentally, these are realizable by preparing the three layers using separate sets of species-selective optical tweezer arrays 
generated by independent spatial light modulators (SLMs)~\cite{Lee2016, Barredo2018, Singh2022}.
Performing the projective measurements on the Cs ancillae in the $X$-basis (equivalently, performing the Hadamard rotations and then measuring in the $Z$-basis), the above proof guarantees that we  obtain the $\ket{\tilde{\XC}}$ state,  related to the true X-cube ground state $\ket{\XC}$ by the local unitaries, as shown above. Importantly, because of the crosstalk-free measurement capabilities in the dual-species architecture, these measurements can be performed without decohering the code qubits and feed-forward operations dependent on the measurement results can be implemented~\cite{Singh2023}.

\vspace{2mm}
\noindent
\textbf{Error-correcting properties of  $\ket{\XC}$}. Given the relatively small depth (3 layers) of the proposed geometry, what properties of the X-code ground state survive? Note that since we do not implement the actual stabilizer Hamiltonian in Eq. (2), it makes it  difficult to talk about the excitations, such as the immobile fractons and particles with subdimensional mobility (lineons, planons). Nevertheless, the ground state $\ket{\XC}$ we engineered possesses nontrivial entanglement between code qubits and is therefore of interest in its own right. 
Can one leverage this entanglement to identify the defects (a.k.a. excitations) in the X-cube state? To answer this question, we consider the error-correcting properties of our framework.

Let us first discuss the errors in the state $\ket{XC}$ we engineered above. Our starting point, the 3D cluster state, has an intrinsic error correction capability (see Ref.~\cite{raussendorf-bravyi-harrington2005}), as opposed to the 1D cluster state~\cite{briegel-raussendorf2001}. Recall that our construction results in essentially the measurements of the $C_c$ and $C_a$ cluster operators in Eq.~\eqref{eq:cluster-ops}.
Although each measurement is individually random, they are not independent -- these measurements must satisfy parity constraints, namely for any code qubit $c$,
\beq
\prod_{a\in\partial c} X_a \prod_{c'\in \partial a} Z_{c'} = \mathbb{1}, 
\label{eq:constraint}
\eeq
and similarly for the ancilla qubits. 
Violation of any of these constraints (syndroms) indicates an error that can be reliably identified and corrected, provided the error rate is not too high~\cite{raussendorf-bravyi-harrington2005}.

\begin{figure}[tbh]
 \centering
  \includegraphics[width=0.40\textwidth]{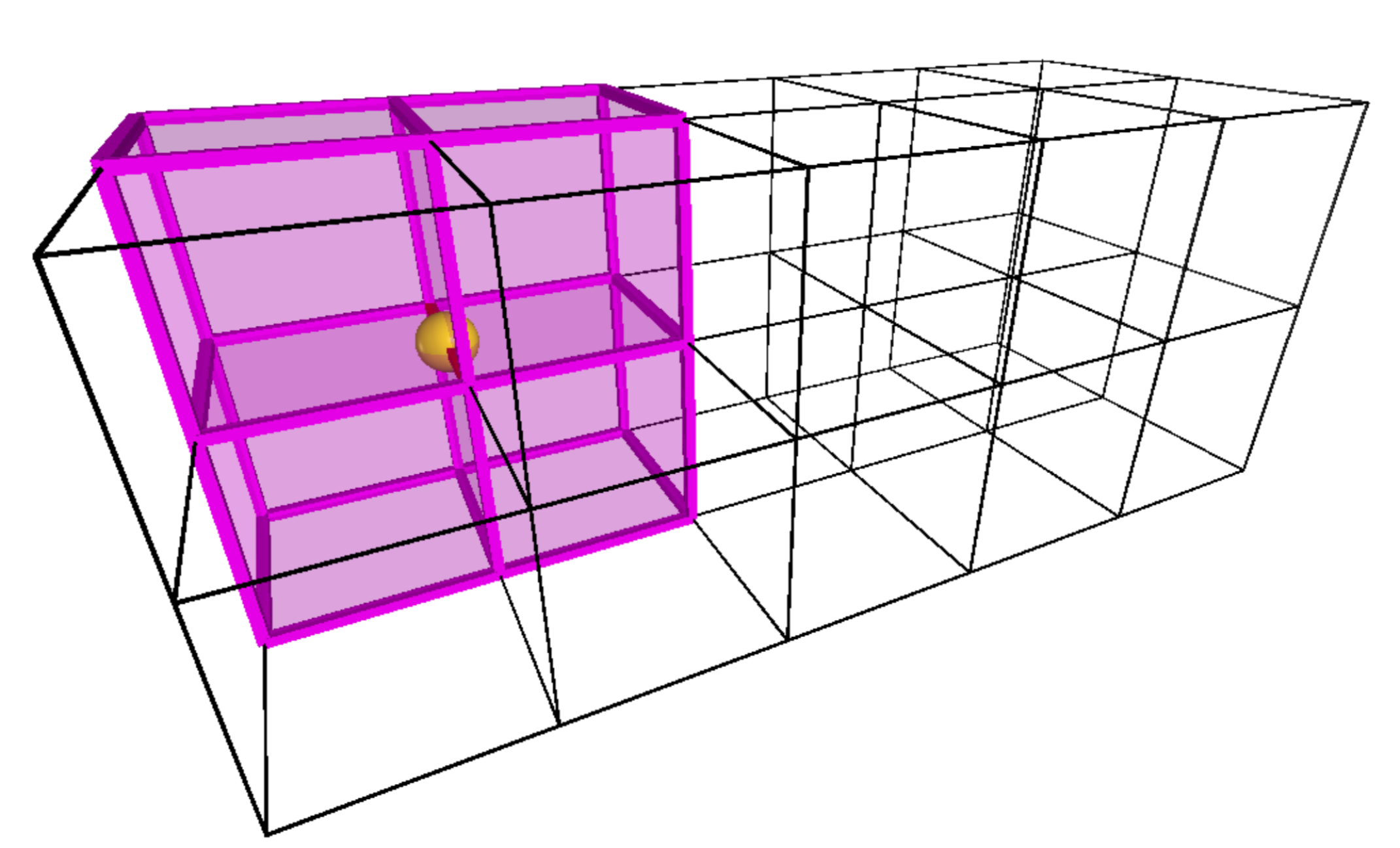}
  \includegraphics[width=0.40\textwidth]{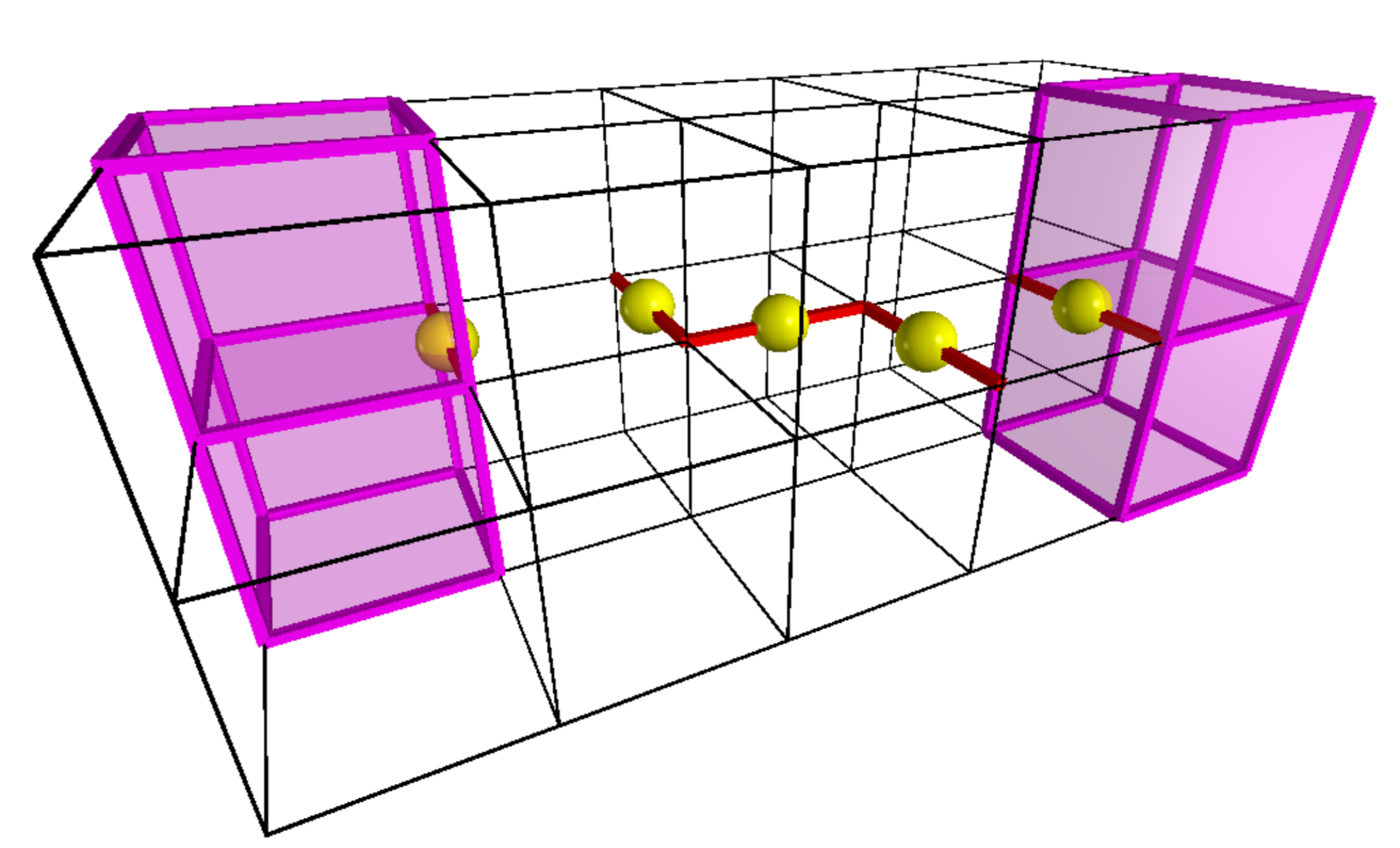}
  \caption{Top: A plaquette of four fractons (magenta cubes) is a result of a single $Z$-error ($X$-operator application, shown with a yellow ball) on a code qubit. Bottom: the motion of a pair of fractons is constrained to be in the plane perpendicular to the dipole orientation, mitigated by a set of $X$ operators along a chain of code qubits.}
  \label{fig:dipole}
\end{figure}

We now consider the errors in the projective measurements of the ancillae that lead to $|\tilde{\XC}\rangle$ -- since those measurements are performed in the $X$-basis, only $Z$-errors (or equivalently, $Y$-errors) affect those. One such error on a given ancilla $a$ creates a syndrome that is reflected in $B_\text{cube}(a)$, and given the constraint \eqref{eq:cube-product}, can be identified. Since an error in one $B_\text{cube}$ is equivalent to creating a fracton excitation in the X-cube model, the above statement indicates that an isolated fracton can be detected.
Let us now consider a dipole of fractons,  that is two nearby $B_\text{cube}$ errors stemming from an $X$ operator on an edge shared by the cubes. The passive error correction encoded in Eq.~\eqref{eq:cube-product} is not sufficient to detect such errors.
Moreover, by acting with a sequence of $X$ operators on nearby edges, the fracton dipole can be made to move in the direction perpendicular to its dipole motion, as shown in Fig.~\ref{fig:dipole}.~\cite{footnote1} 

\begin{figure}[tbh]
  \centering
  \includegraphics[width=0.40\textwidth]{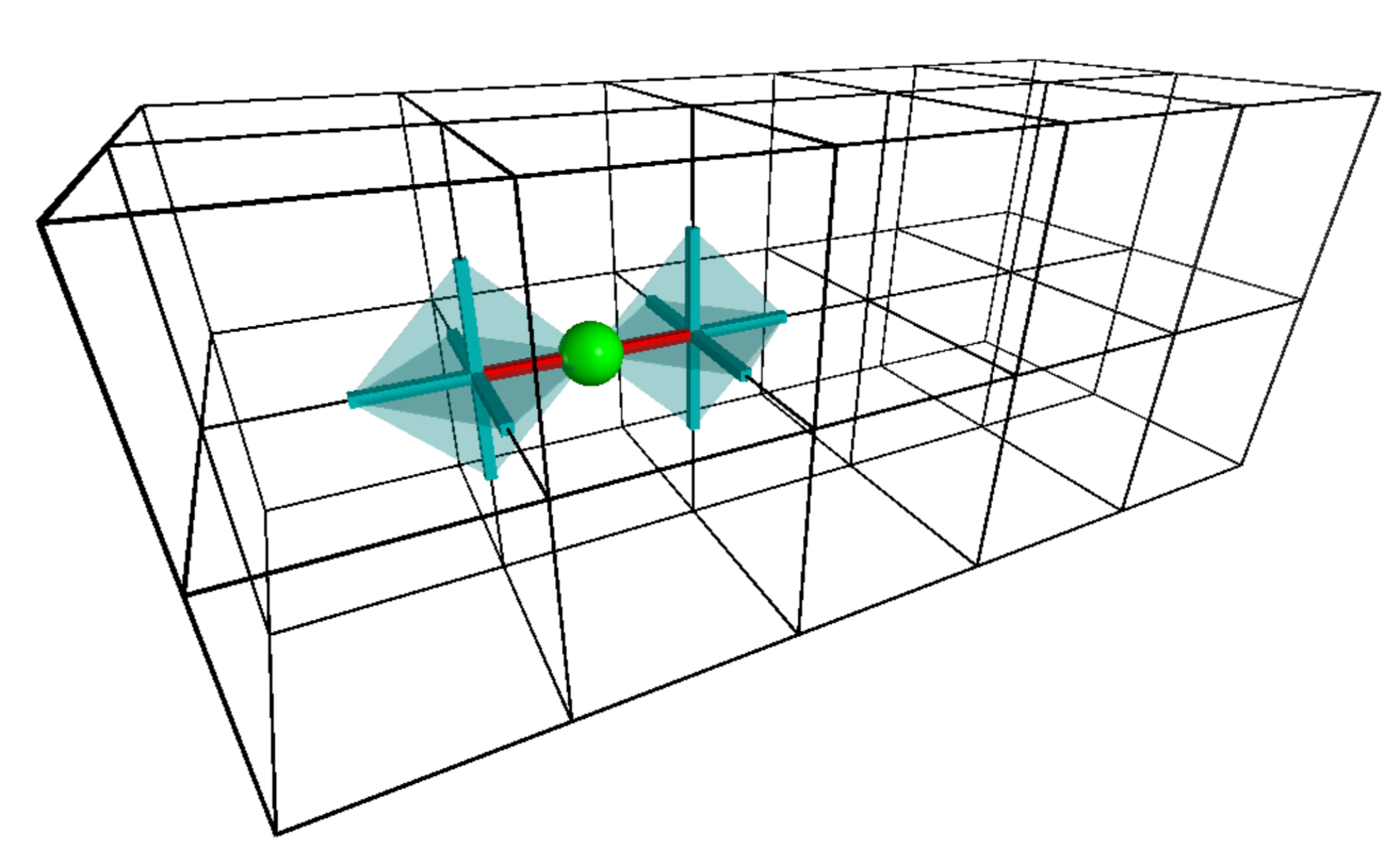}
  \includegraphics[width=0.40\textwidth]{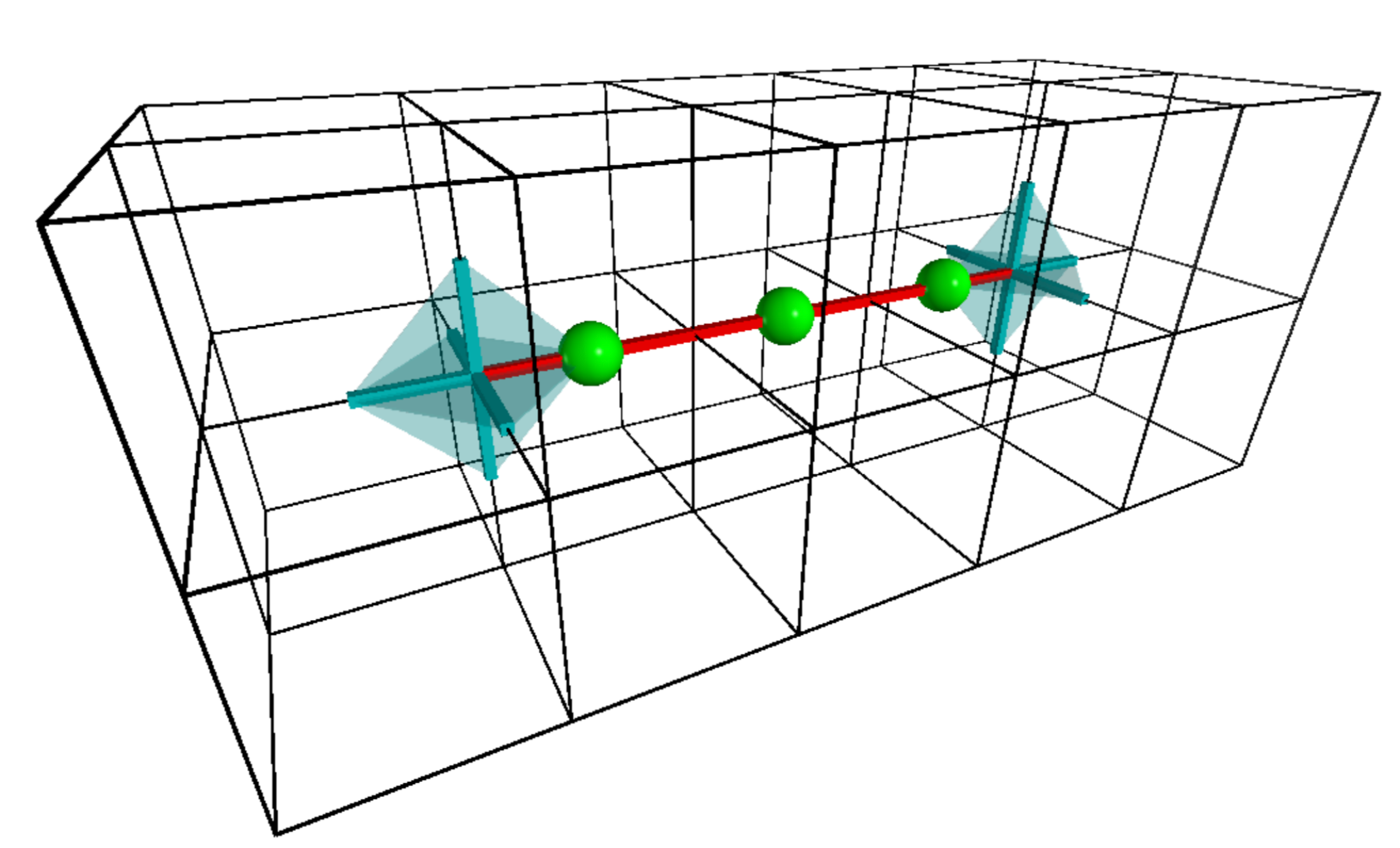}
  
  \caption{Top: a pair of lineon excitations (each shown as two cyan crosses) is formed by acting with a $Z$ operator on the green qubit. Bottom: lineons can only move along 1D direction by acting with successive $Z$ operators on the (green) qubits along the line.}
  \label{fig:lineon}
\end{figure}

Finally, we consider an $X$-error (application of a $Z$ operator) on a code qubit. Such an error does not affect the $B_\text{cube}$ terms, but instead creates a pair of defects/syndroms on the nearby star terms in Eq.~\eqref{eq:Xcube}. In the context of the X-cube model, this is a lineon excitation, which can travel in one direction by appending further $Z$-operators along the line, see Fig.~\ref{fig:lineon}. In our implementation, we do not measure the syndroms on the star stabilizers directly, but instead, an $X$ error on the code qubit can be detected by measuring the qubit-centered $C_c$ cluster operator in Eq.~\eqref{eq:cluster-ops} that includes 4 nearby ancillae -- the presence of an error will be reflected in the violation of the parity constraint (similar to Eq.~\ref{eq:constraint})
\beq
\prod_{c} X_c  \prod_a Z_a = \mathbb{1}.
\eeq

\textit{En pr\'ecis,} We have demonstrated how the entanglement resource stored in the cluster state involving the code and ancilla qubits can be leveraged to create a ground state of the X-cube model. We propose to utilize the recently developed dual-species (Rb and Cs) capability of programmable tweezer arrays of  Rydberg atoms as a concrete platform. Instead of directly implementing the stabilizer terms of the X-cube model, which is challenging because of the multi-atom nature of the necessary interactions, our program requires only pairwise control-phase gates between the ancilla and code qubits, whose implementation is readily available using the Rydberg tweezer arrays. \textit{En route} to this goal, we show that our approach, using measurements on the ancilla qubits, could be cast in the language of so-called  dynamic quantum circuis with mid-circuit measurements, which have been advertised as an efficient means of creating long-range entanglement using only shallow-depth circuits.
We further demonstrate that our scheme -- although not routed in the Hamiltonian approach, and therefore not readily designed to address the excited states of the X-cube code -- nevertheless allows us to detect and correct certain errors that correspond to the fracton and lineon excitations of the X-cube model. While a true fracton order would require a three-dimensional lattice, our simplified ``one-storey'' model consists of three two-dimensional layers for reasons of practicality of its implementation in the Rydberg tweezer platform. Nevertheless, we view this as an intermediate step \textit{en route} to engineering the true fracton order. 

\renewcommand{\emph}[1]{\textit{#1}}
\bibliographystyle{nature}
\bibliography{fractons,Rydberg,cluster,footnotes,HB-refs}

\vspace{5mm}
\noindent
\textbf{Methods}

\vspace{1mm}
\noindent
{\small\textbf{Details of the dual-species Rydberg tweezer platform}\vspace{2mm}

\noindent
The proposed protocol can be conveniently realized in a dual-species Rydberg array as developed by one of us~\cite{anand_dual-species2024}. Here, the two species can be independently controlled and measured. For concreteness, the two atomic species can be individually trapped rubidium (Rb) atoms and cesium (Cs) atoms as code and ancilla qubits respectively. The qubits are encoded in the $\ket{0}$ and $\ket{1}$ hyperfine states of the respective atoms, with single-qubit operations performed via the microwave or Raman driving, see Fig.~\ref{fig:cluster}(a). The Rydberg blockade is achieved via application of the species-selective $\ket{1}\to \ket{r}$ ground-Rydberg laser. Since the laser frequencies that address these transitions are far-detuned from one another, the principal quantum number for the Rydberg states $\ket{r}_\text{Rb}$ and $\ket{r}_\text{Cs}$  can be chosen independently for the two species. As demonstrated in an earlier work by one of us~\cite{anand_dual-species2024}, Rb-Cs pairs excited to $\ket{68 S_{1/2}}_\text{Rb} \ket{67S_{1/2}}_\text{Cs}$ state  undergo resonantly enhanced  F\"orster interactions (due to near-degeneracy with the neighboring pair state $\ket{67P_{1/2}}_\text{Rb} \ket{67P_{3/2}}_\text{Cs}$), enabling a tunable interspecies interaction, much stronger than the intra-species one. This asymmetric interaction regime allows for the implementation of the CZ$_{12}$ gate without movement of the atoms, as described in the maintext.

}

\vspace{2mm}
\noindent
{\small\textbf{Two-qubit control-phase gates with Rydberg atoms}

\vspace{1mm}

\noindent
The aforementioned tunable F\"orster interaction between the two species paves the way to an effective implementation of the interspecies two-qubit gates involving a code qubit (Rb) and an ancilla (Cs). In particular, we demonstrated~\cite{anand_dual-species2024} that an interspecies CZ gate can be implemented by the $\pi$-$2\pi$-$\pi$ sequence~\cite{Jaksch2000}, see Fig.~\ref{fig:cluster}(b).

As discussed in the main text, we envision two approaches to constructing the cluster state. The first one relies on  moving the positions of the ancilla (Cs) atoms relative to the Rb sublattice in the direction $R_i$ ($i=1..12$), as depicted by the green arrows in Fig.~\ref{fig:cluster}c). This procedure can be performed in parallel on all the periodic images of the $i^\text{th}$ code qubit, thus requiring a total of 12 operations. 

\begin{figure}[tbh]
  \centering
  \includegraphics[width=0.45\textwidth]{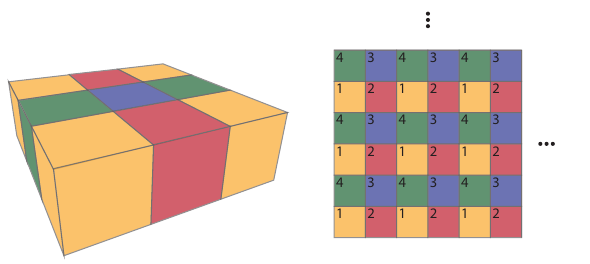}
  \caption{Schematic of applying a series of CZ$_{12}$ gates. Each colored cube represents an ancilla connected via CZ$_{12}$ gate to the surrounding 12 qubits, as shown in Fig. 2c in the main text. The Rydberg blockade means that the gates must be applied sequentially on differently colored cubes: (1) all yellow cubes, (2) all red cubes etc. Within a given color, the gates can be applied in parallel (i.e. all yellow cubes at the same time, etc).}
  \label{fig:seq-CZ12}
\end{figure}

\vspace{3mm}
\noindent
{\small\textbf{CZ$_{12}$ control-phase gates with Rydberg atoms}

\vspace{1mm}
\noindent
The method used to apply a CZ gate to a pair of Rydberg atoms can be generalized to a CZ$_n$ gate, which conditionally applies a Z gate to multiple $n$ target qubits neighboring the ancilla. Importantly, to realize this gate, only the ancilla qubit and the neighbors should experience Rydberg blockade, while the neighbors among themselves should not be blockaded. In this regime, sending a $\pi$ pulse to the ancilla qubit, a $2\pi$ pulse to each neighboring code qubit, and another $\pi$ pulse to the ancilla applies the CZ$_n$ gate. Applying this sequence to an elementary cube, realizes the desired connectivity as shown schematically in Fig.~\ref{fig:cluster}(b). 
This is equivalent to twelve CZ gates, without the need to move the ancilla atom twelve times.

Thus, our second approach consists of implementing CZ$_{12}$ gates between a given ancilla and the surrounding code qubits. The subtlety is that the CZ$_{12}$ gates can only be performed simultaneously on non-overlapping cubes (i.e. cubes that do not share any surfaces, edges or corners). As the Fig.~\ref{fig:seq-CZ12} illustrates, this can be achieved in four steps, each one acting in parallel on the cubes of a given color.
}

\vspace{2mm}
\noindent
{\small\textbf{Details of cluster state preparation}

\vspace{1mm}
\noindent
Below, we provide detailed proof that applying a sequence of pairwise CZ gates to the ancilla and code qubits prepared in the $\ket{+}$ state, as shown in the quantum circuit in Fig.~3(c) in the main text,  indeed results in a cluster state.

Consider two qubits $a$ and $c$ prepared in the $\ket{+}$ state and entangled via a $CZ$ gate. Knowing that $(CZ)_{ac} X_a (CZ)_{ac}=X_a Z_c$, it follows that the entangled state $(CZ)_{ac}\ket{+_a+_c}$ is the $+1$ eigenstate of the cluster operators $X_aZ_c$ and $X_cZ_a$. This can be extended to prove that the cluster state of any connected graph of qubits can be created by preparing all qubits in the $\ket{+}$ state and applying a $CZ$ gate to every pair of connected ancilla-code qubits. In the case of our cluster state used to prepare the X-cube code, twelve $CZ$ gates are needed for each ancilla qubit, as each one has $n=12$ neighbors (see Fig.~\ref{fig:cluster}(c)). The resulting state is then a $+1$ eigenstate of the cluster operators 
\begin{equation*}
C_c = X_c \prod_{a\in \partial c} Z_a, \quad
C_a = X_a \prod_{c\in \partial a} Z_c. 
\end{equation*} 
\begin{flushright}
    \vspace{-3mm}
    {Q.E.D.} 
\end{flushright}

}

\vspace{2mm}
\noindent
{\small\textbf{Dynamical quantum circuit}

\vspace{1mm}
\noindent
In Fig. \ref{fig:circuit}, the circuit was drawn to most directly reflect the procedure performed in a programmable Rydberg tweezer array to prepare the code state. However, note that the CZ gate is equivalent to a CNOT gate preceded and followed by a Hadamard gate applied to the qubit on which the NOT is performed. Therefore, the circuit can be redrawn as shown in Fig. \ref{fig:circuit-CX}, with the initial state and measurements both performed in the $Z$-basis. The resulting circuit is equivalent to the so-called \textit{dynamical} GHZ-state preparation circuit proposed in Ref. \cite{baumer_dynamic-circuits2023} modulo the classical corrections applied afterwards to correct for single-qubit phases. This shows that there is a straightforward way to perform our method of X-cube code state preparation on a quantum computer, along with any other code which can be prepared through cluster state projection.

\begin{figure}[tbh]
  \centering
  \includegraphics[width=0.33\textwidth]{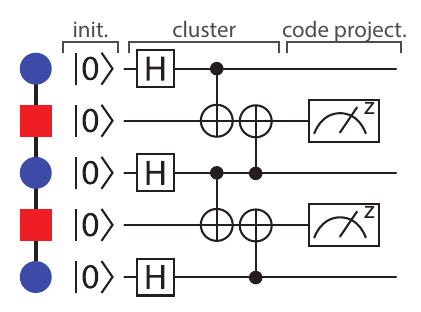}
  
  \caption{The schematic of the quantum circuit using CNOT operators, equivalent to the circuit shown in Fig.~\ref{fig:circuit}(c). As before, this realizes an eigenstate of the X-cube model.
  }
  \label{fig:circuit-CX}

\end{figure}

}



\vspace{2mm}
\noindent
\textbf{Data availability}
\newline\noindent
{\small The supporting data are available from the corresponding author upon reasonable request.
}

\vspace{2mm}
\noindent
\textbf{Acknowledgements}\newline\noindent
{\small
A.H.N. was supported by the Department of Energy under the Basic Energy Sciences award no\ldots. H.B. thanks Hannes Pichler for insightful discussions. H.B. acknowledges funding from the Office of Naval Research (N00014-23-1-209 2540) and the Air Force Office of Scientific Research (FA9550-21-1-0209). A.H.N. and H.B. acknowledge the hospitality of the Aspen Center for Physics, supported by the National Science Foundation grant PHY-1607611.
}\newline\newline

\vspace{2mm}
\noindent
\textbf{Author contributions}\newline\noindent
{\small
A.H.N. designed the project and performed the theoretical analysis with the help of A.C.  H.B. contributed the details of the dual-species Rydberg tweezer arrays. All authors contributed to the discussion of the results. A.H.N. wrote the manuscript with contributions of all the authors.
}\newline\newline

\vspace{2mm}
\noindent
\textbf{Competing Interests}\newline\noindent
{\small
The authors declare no competing interests.
}\newline\newline

\vspace{2mm}
\noindent
\textbf{Additional Information}\newline\noindent
{\small
Correspondence should be addressed to A.H.N.
}

\end{document}